\documentclass[times,twocolumn,final]{elsarticle}
\usepackage{outline}
\usepackage{pmgraph}
\usepackage[normalem]{ulem}
\usepackage[utf8]{inputenc}
\usepackage{amssymb}
\usepackage{hyperref}
\usepackage{amsmath}
\usepackage{graphicx}
\usepackage{times}
\usepackage{xcolor}
\usepackage{xspace}
\usepackage[colorinlistoftodos]{todonotes} % for \todo
\usepackage{bm} % bold
\usepackage{url}

\usepackage{epstopdf}
\AppendGraphicsExtensions{.tiff}
\graphicspath{{fig/}} % Directory in which figures are stored

\usepackage{epsfig}
\usepackage{tikz}
\usetikzlibrary{spy}
\usepackage{algpseudocode}
\usepackage{algorithm}
\usepackage{mathrsfs}

\def\QED{~\rule[-1pt]{5pt}{5pt}\par\medskip}

\long\def\comment#1{} % comment out text

\newcommand{\xmath}[1] {\ensuremath{#1}\xspace}
\newcommand{\blmath}[1] {\xmath{\bm{#1}}}

\newcommand{\x}{\blmath{x}}

\newcommand{\beq}{\begin{equation}}
\newcommand{\eeq}{\end{equation}}
\newcommand{\beqa}{\begin{eqnarray}}
\newcommand{\eeqa}{\end{eqnarray}}

\usepackage[mathlines]{lineno}

\usepackage{multirow}
\usepackage{makecell}
\usepackage{booktabs}
\usepackage[flushleft]{threeparttable}
\usepackage{etoolbox}
\usepackage{amsmath,soul}
\AtBeginEnvironment{figure}{\setlength{\intextsep}{0.5\baselineskip}}
\usepackage[algo2e]{algorithm2e}
\SetKwInput{KwInput}{Input}
\SetKwInput{KwOutput}{Output}

\usepackage{medima}
\usepackage{framed,multirow}

\usepackage{latexsym}

\usepackage{url}
\usepackage{xcolor}
\usepackage{hyperref}

\definecolor{newcolor}{rgb}{.8,.349,.1}

\begin{document}

\verso{J. Huh \textit{et~al.}}

\begin{frontmatter}
\title{Breast Ultrasound Report Generation using LangChain}

\author[1]{Jaeyoung Huh}
\author[2]{Hyun Jeong Park\corref{cor1}}
\author[3]{Jong Chul Ye\corref{cor1}}
\cortext[cor1]{Corresponding authors:  E-mail address : seolly1024@cau.ac.kr (H.J.Park); jong.ye@kaist.ac.kr (J.C.Ye)}

\address[1]{Department of Bio and Brain Engineering, Korea Advanced Institute of Science and Technology (KAIST), Daejeon 34141, Republic of Korea}
\address[2]{Department of Radiology, Chung-Ang University Hospital, Chung-Ang University College of Medicine, Heukseok-ro, Dongjak-gu, Republic of Korea}
\address[3]{Kim Jaechul Graduate School of AI, Korea Advanced Institute of Science and Technology (KAIST), Daejeon 34141, Republic of Korea}

\begin{abstract}
Breast ultrasound (BUS) is a critical diagnostic tool in the field of breast imaging, aiding in the early detection and characterization of breast abnormalities. Interpreting breast ultrasound images commonly involves creating comprehensive medical reports, containing vital information to promptly assess the patient's condition. However, the ultrasound imaging system necessitates capturing multiple images of various parts to compile a single report, presenting a time-consuming challenge. To address this problem, we propose the integration of multiple image analysis tools through a LangChain using Large Language Models (LLM), into the breast reporting process. Through a combination of designated tools and text generation through LangChain,
our method can accurately extract relevant features from ultrasound images, interpret them in a clinical context, and produce comprehensive and standardized reports. This approach not only reduces the burden on radiologists and healthcare professionals but also enhances the consistency and quality of reports. The extensive experiments shows that each tools involved in the proposed method can offer qualitatively and quantitatively significant results. Furthermore, clinical evaluation on the generated reports demonstrates  that the proposed method can make report in clinically meaningful way.
\end{abstract}

\end{frontmatter}

 \begin{figure*}[t]
  \center
	\includegraphics[width=\textwidth]{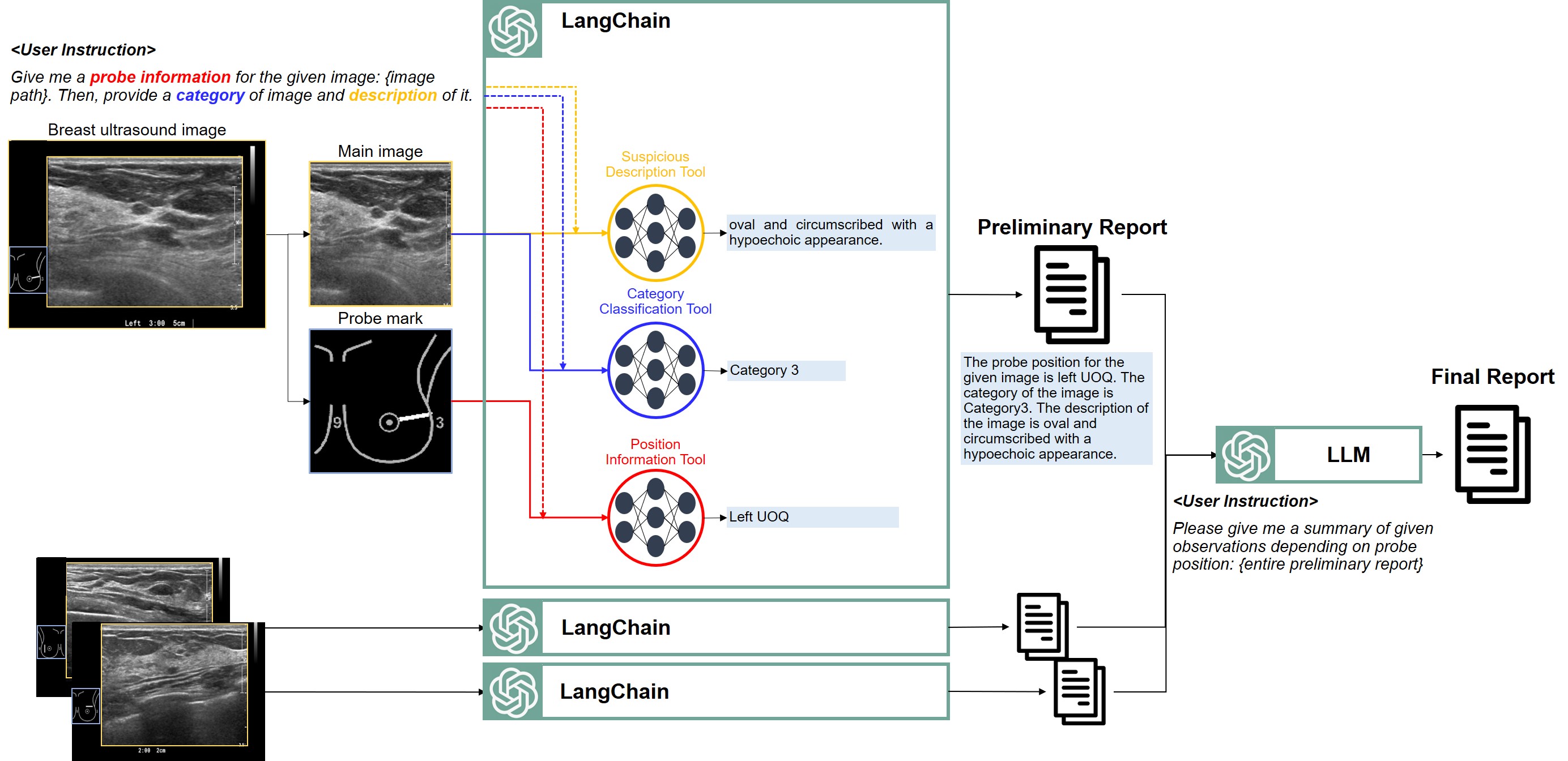}
	\caption{Overal framework of the proposed method. The process involves inputting BUS images into LangChain, which is equipped with specialized tools such as the `Suspicious description tool', `Category classification tool,' and `Probe information tool,' guided by user instructions. Subsequently, LangChain generates a preliminary report for each image. These individual preliminary reports can then be summarized using LLM, ultimately generating the final report.}
	\label{fig:framework}
\end{figure*}

\section{Introduction}
\label{sec:introduction}
Breast ultrasound (BUS) serves as a rapid and straightforward method for the initial diagnosis of breast abnormalities. Achieving precise interpretations of BUS images is vital for guiding clinical decisions, yet this often entails the laborious process of manually crafting comprehensive medical reports. To generate a  single report for BUS,
radiologists  should capture multiple BUS images  at specific breast reference points or suspicious scenes during the scanning. Later captured frames are referenced in generating a single report.
Since these reports have been the domain of radiologists and healthcare experts, it demanding a profound comprehension of breast pathology and substantial time commitment.
%Complicating matters further, each radiologist employs their unique judgment criteria, resulting in inconsistency and a decline in report quality.

%Traditionally, generating medical reports automatically has often hinged on structured templates and rule-based systems to organize data within predefined guidelines. This method heavily depends on fixed templates and rules, which limits its adaptability when dealing with nuanced cases and its ability to capture intricate clinical insights.
 In recent years, the emergence of deep learning techniques, particularly in the field of Natural Language Processing (NLP), has significantly enhanced workflows in the medical field. These advancements have opened doors to the creation of automated systems capable of aiding in medical image analysis and report generation.

For example, the authors  of \citep{zhang2017mdnet,jing2018automatic,yuan2019automatic,gale2019producing,yang2021automatic} presented a method for generating diagnostic reports by employing a neural network that combines Convolutional Neural Network (CNN) and Long Short-Term Memory (LSTM). Similarly, the authors of \citep{li2018hybrid,li2019knowledge,xu2020reinforced} introduced a medical image report generation method based on reinforcement learning and graph neural networks. The Vision Language Pre-trained (VLP) model has rapidly advanced in medical applications, acting as a versatile pre-trained model enabling accurate and efficient diagnosis by leveraging both visual and textual information \citep{yan2022clinical,moon2022multi,lee2023unified}.

% In particular, the author of \citep{yan2022clinical} proposed 'Clinical-BERT' model which is based on the VLP architecture and can be harnessed as an expert in downstream tasks such as report generation and diagnosis. Furthermore, the author of \citep{moon2022multi} put forward a method based on transformer architecture with a self-attention mechanism, demonstrating superior results in various down stream tasks including VQA, image-report retrieval, and report generation. Lastly, the author of \citep{lee2023unified} proposed a method for bi-directional Chest X-ray (CXR) and report generation, employing vector quantization method. 

The Large Language Model (LLM) is a foundational model trained on a vast corpus of text, enabling it to grasp the intricacies of grammar, syntax, semantics, and even world knowledge. Thanks to its versatility, numerous researchers have sought to harness the power of the LLM for their specific needs, such as in the field of medical knowledge \citep{singhal2023large,moor2023foundation,nori2023capabilities}. Furthermore, a multimodal approach involving the LLM has been extensively studied to attain expert-level clinical applications \citep{thawkar2023xraygpt,moor2023med,xu2023elixr}. Not only for language tasks, the LLM has been utilized as powerful agents supporting to achieve specific purpose. Beyond language-related tasks, the LLM serves as a potent assistant, capable of working semi-autonomously in a variety of applications, from conversational chatbots to goal-driven workflow automation. 

Moreover, the LLM can be equipped with an array of external tools, including open-source utilities, and schedule the entire process using designated tools to achieve specific objectives \citep{qin2023toolllm,parisi2022talm,schick2023toolformer,shen2023hugginggpt}. Among the available tools, LangChain, an open-source framework built on the LLM, facilitates the optimal selection of tools to accomplish predefined goals \footnote{https://github.com/langchain-ai/langchain}. The LLM's decision-making is context-driven, determining how to respond and which actions to take based on the provided context.

Building upon the pioneering works that have inspired us, we present a novel approach for generating BUS reports, leveraging the LangChain framework as illustrated in Figure 
 \ref{fig:framework}. This method is crafted from a collection of specialized tools, each excelling in its respective task, thereby enhancing result accuracy. Our approach retains the inherent capabilities of the LLM, including a conversational interface, while additionally performing BUS analysis functions through the careful selection of tools. Notably, the memory system within LangChain empowers the method to recapitulate prior observations, facilitating the generation of preliminary reports. 
 
 In contrast to previous methods, which could only summarize the central image elements, our proposed approach excels in recognizing peripheral areas of the image containing crucial probe information essential for report generation. Furthermore, our method comprises straightforward networks such as ResNET-50 \citep{he2016deep}, ensuring ease of training and, consequently, straightforward upgrades. Our experimental results support the effectiveness of each designated tool, emphasizing their specialization in their respective tasks and affirm that our proposed method significantly success BUS report generation.
 
Our main contributions can be summarized as follows:
\begin{itemize}
    \item We propose a method for generating BUS reports, achieved through the combination of specialized tools within the LangChain framework.
    \item These tools are fashioned with uncomplicated classification networks, which facilitates easy training and upgradability or expansion.
    \item Preliminary reports generated from various breast parts using LangChain's memory system can be transformed into a final report using another LLM.
\end{itemize}   

This paper is structured as follows. First, we offer an introduction to the LLM as an agent and its relevance in medical report generation. Subsequently, we delve into the specifics of our proposed method in Section \ref{sec:contribution}, followed by Section \ref{sec:implementation}. Next, we present the results of the proposed method and engage in a comprehensive discussion in Section \ref{sec:results} and Section \ref{sec:Discussion}, respectively.

 \begin{figure*}[t]
  \center
	\includegraphics[width=0.75\textwidth]{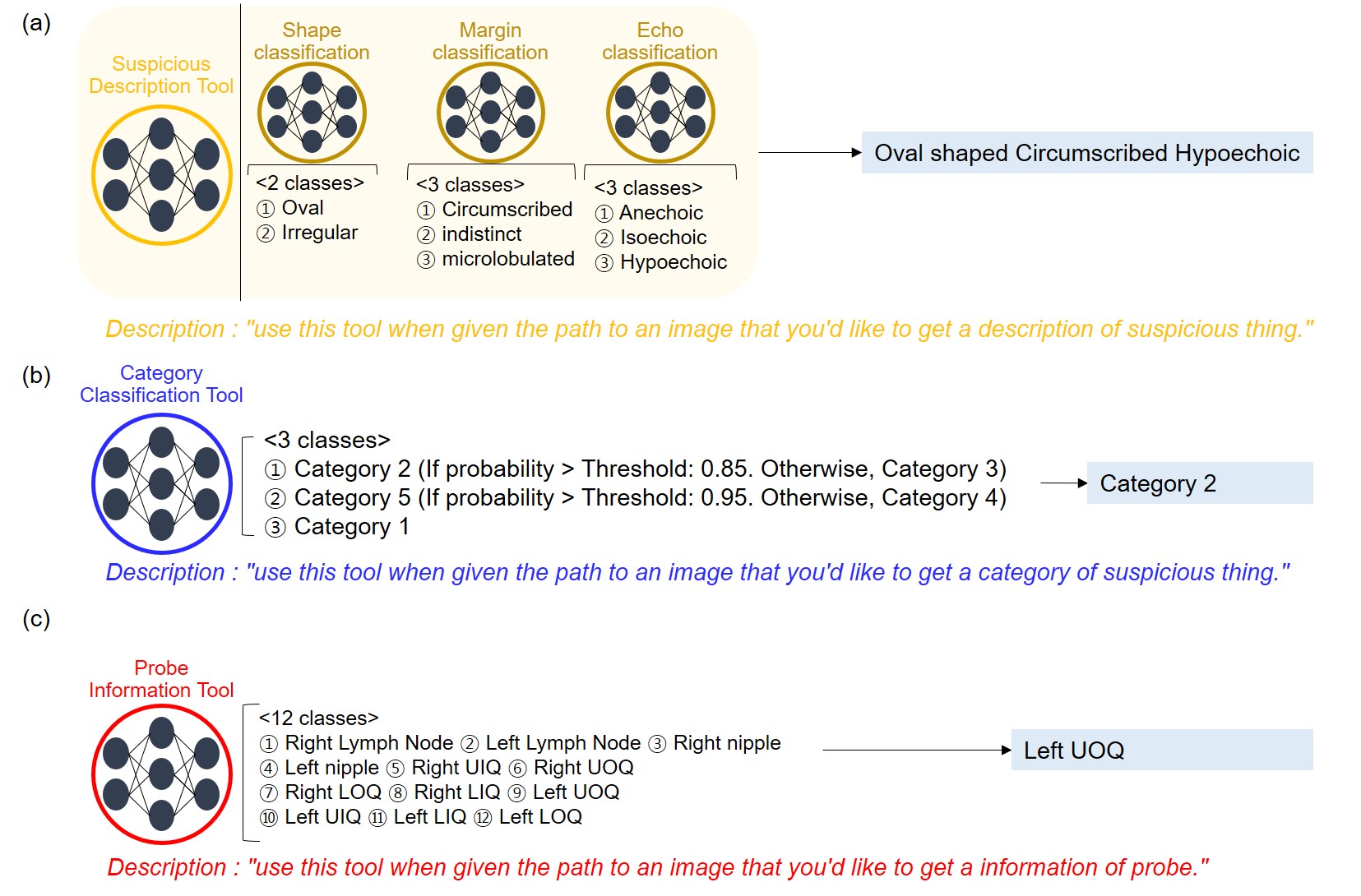}
	\caption{Details of tools. The description provides an explanation of the tools utilized within the LangChain framework. (a) The `Suspicious description tool' combines shape, margin, and echo classification networks. (b) The `Category classification tool' classifies into 3 distinct classes. (c) The `Probe information tool' categorizes into 12 specific classes.}
	\label{fig:network}
\end{figure*}

\section{Background}
\label{sec:background}

\subsection{LLM as agent}
The LLM, a neural network trained on extensive text corpora \citep{raffel2020exploring,brown2020language,ouyang2022training,touvron2023llama}, possesses an exceptional capacity to comprehend, generate, and manipulate human language. This profound understanding of natural language renders it a valuable tool. Nevertheless, the LLM does have a few limitations. Its primary input and output are text, which makes it challenging to apply in multi-modal tasks. Additionally, the LLM may exhibit performance degradation during fine-tuning when the dataset is insufficient or of sub-optimal quality.
Moreover, in certain scenarios, several sub-tasks are necessary to achieve a singular purpose, often requiring a sequential process. 

In response to this challenge, several research endeavors have explored harnessing the LLM as an agent to oversee the management of tools. The author of \citep{shen2023hugginggpt} introduced a method employing the LLM as a controller for scheduling tasks through steps encompassing task planning, model selection, task execution, and response generation. This approach notably combines ChatGPT with the resources provided by Hugging Face, a prominent open-source community offering a wide array of pre-trained models. The author of \citep{schick2023toolformer} presented a method focused on determining which APIs to invoke, utilizing GPT-J \footnote{https://github.com/kingoflolz/mesh-transformer-jax} as a foundational element in their strategy. Furthermore, LangChain, as detailed in its repository \footnote{https://github.com/langchain-ai/langchain}, serves as a framework meticulously crafted to facilitate the creation of applications leveraging large-scale language models. As a comprehensive language model integration framework, LangChain's applications align closely with those commonly associated with language models, encompassing document analysis, summarization, chatbots, and code analysis, as well as the ability to incorporate user-defined tools tailored to their specific needs. Moreover, harnessing its memory system, LangChain has the capacity to implement instructions based on prior memory, enhancing its adaptability and utility.

\subsection{Medical report generation}
Medical report generation is imperative for enhancing workflow within the medical field. The manual creation of reports from medical images is a time-consuming process, and the results can vary depending on the clinician. Numerous research efforts have been dedicated to improving medical report generation, with deep learning-based approaches showing remarkable performance. In the early stages, a combination of CNN and RNN or LSTM architectures, often coupled with attention mechanisms, was prevalent \citep{shin2015interleaved,shin2016learning,jing2017automatic}. 

The emergence of transformer architectures has led to significant advancements in the field of NLP \citep{boecking2022making,huang2021gloria}. The author of \citep{liu2021exploring} introduced a method for automating the generation of radiology reports, employing a transformer encoder-decoder architecture with the Posterior-and-Prior Knowledge Exploring-and-Distilling (PPKED) approach. Similarly, the author of \citep{wang2021self} put forth a self-boosting framework designed to foster cooperative learning between image-text matching and progressive report generation. Notably, the VLP model, which integrates vision and text information, has achieved remarkable performance across various NLP tasks, including Vision VQA, diagnosis classification, image-report retrieval, and report generation \citep{moon2022multi,bannur2023learning}.

The emergence of Large Language Models (LLM) has spurred rapid research in the field of report generation and diagnosis. The author of \citep{wang2023chatcad} harnessed LLM as a tool to distill concise reports from various outputs of multiple Computer-Aided Diagnosis (CAD) systems. The author of \citep{zhou2023skingpt} introduced a method for evaluating skin images, employing fine-tuning with MiniGPT-4 \citep{zhu2023minigpt}. Furthermore, the author of \citep{zhong2023chatradio} presented an approach for generating generalizable radiology reports, bridging cross-source heterogeneity issues using LLM.

The majority of research in the domain of report generation has predominantly focused on CXR datasets, with relatively limited attention given to ultrasound report generation. In one notable effort, the author of \citep{yang2021automatic}, introduced a method for generating breast ultrasound reports, employing a combination of CNN and LSTM architecture. This approach excelled at describing image attributes such as echoes and shapes, but was unable to provide positional information as it solely utilized the ultrasound image component. Furthermore, the author of \citep{ge2023ai} proposed a method for generating screening reports for BUS images, capitalizing on a range of classification networks combined with a fusion of CNN and transformer architectures.

\section{Main Contribution}
\label{sec:contribution}

\subsection{LangChain-based BUS report generation}
LangChain, a framework rooted in LLM, facilitates the development of Artificial Intelligence (AI) applications with remarkable ease, allowing for the easily combination of various models. While the original LLM primarily operates with textual data and cannot handle multi-modal datasets, its agent, LangChain, extends its capabilities to process diverse data types, including images, by invoking specialized tools. In particular, the LangChain agent can automatically access tools via text instructions. It is equipped to determine which tool to employ based on user instructions, execute the tool, and provide the output in text format. For instance, when a user provides an instruction specifying an image path, such as ``What is the probe information of the given image: $\{\textit{image path}\}$," the LLM determines the need for the probe position information tool and generates a response like, ``The probe information of the given image is right Lower Outer Quadrant (LOQ)." The LLM selects the appropriate tool based on the description of each designated tool.

One of the most crucial characteristics of LangChain is its exceptional memorization ability and, enabling it to store past interaction information. When a user instruction is provided, LangChain engages in the core process using information retrieved from its memory system. Furthermore, LangChain facilitates the development of thought chains that dictate the sequence of tasks. For instance, when a user requests, ``Give me a probe information for the given image: $\{\textit{image path}\}$. Then, provide a category of image and description of it." the LangChain tool effectively processes and delivers the desired results in a sequential manner. In the final step, it provides a summary of the observations, which we refer to as a preliminary report, as outlined in Algorithm \ref{alg:preliminary}. Ultimately, the generation of the final report involves processing the complete preliminary report through the LLM. This process entails an instruction such as ``Please provide a summary of the given observations based on a probe position: \textit{\{entire preliminary report\}}." Notably, in this study, we employed ChatGPT-3.5 \footnote{https://chat.openai.com/chat} as the base LLM.

\begin{algorithm}[H]
 \KwInput{Image path}
 \KwOutput{Preliminary report}
 The user provides an instruction to extract information.\\
 \While{not end of chain-of-thought}{
    The LLM determines the appropriate tool. \\
    The tool is implemented. \\
    Print results \\
  \eIf{end of chain-of-thought}{
       finish loop
   }{
   go back to the beginning of current loop
  }
 }
Print summary of previous observations.
 \caption{How to generate preliminary report}
 \label{alg:preliminary}
\end{algorithm}

Analyzing a single BUS image necessitates the utilization of multiple tools. As depicted in Figure \ref{fig:framework}, the original BUS image encompasses various information, including probe position and main image details. To effectively address this diverse range of information, we have introduced a method that combines several specialized neural networks using LangChain. 
More specifically, we have designated these three tools: the `Suspicious description tool', the `Category classification tool', and the `Probe information tool'. In the subsequent section, we will provide in-depth explanations of each of these tools.

\begin{figure*}[t!]
  \center
	\includegraphics[width=0.8\textwidth]{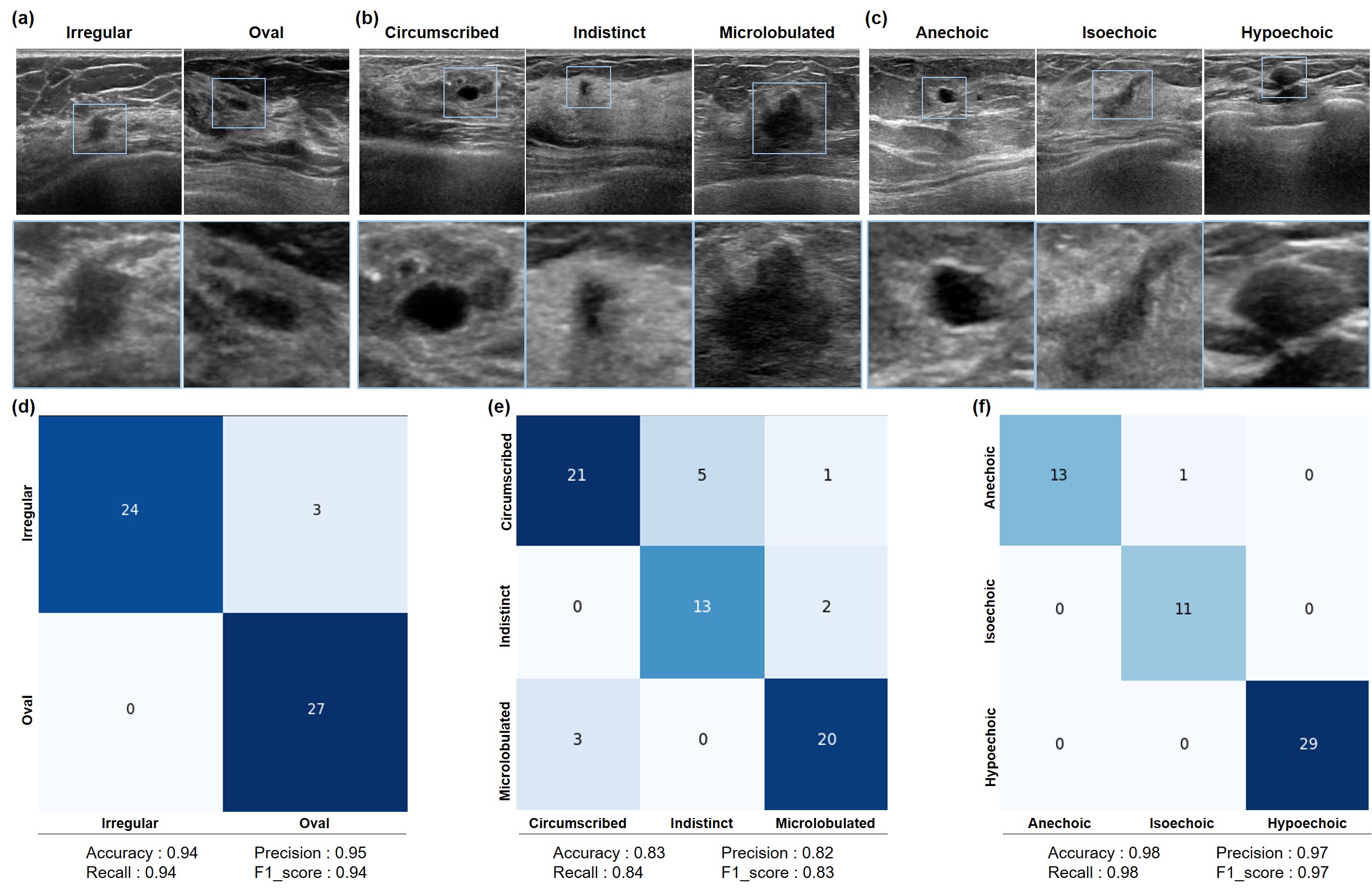}
	\vspace*{-0.3cm}
	\caption{Results of the proposed `Suspicious description tool'. Each number in the confusion matrix denotes the number of each case. (a) The representative results of shape classification. (b) The representative results of margin classification. (c) The representative results of echo classification. (d) The confusion matrix of shape classification. (e) The confusion matrix of margin classification. (f) The confusion matrix of echo classification.}
	\label{fig:result_description}
\end{figure*}

\subsection{Details of tools}
As previously mentioned, we incorporated three tools primarily built around classification models. We independently trained each of these models. It's worth noting that our approach aimed to leverage models that were already established and user-friendly to demonstrate that our proposed method can be effectively implemented with minimal effort.

As depicted in Figure \ref{fig:network} (a), the `Suspicious description tool' comprises three classification networks. These networks collectively describe the characteristics of the suspicious entity, including its shape, margin, and echo. The shape classification is tasked with discerning whether the entity is oval or irregular, while the margin classification classifies it as circumscribed, indistinct, or microlobulated. Furthermore, the echo classification categorizes the entity as anechoic, isoechoic, or hypoechoic. Notably, each of these networks is built upon the pre-trained ResNET-50 model \citep{he2016deep}.

In Figure \ref{fig:network} (b), the `Category classification tool' is displayed, responsible for categorizing images into three main groups according to Breast Imaging-Reporting and Data System (BI-RADS): negative (C1), benign (C2), or highly suggestive of malignancy (C5). If the network's output, representing the likelihood of belonging to each of these categories, falls below a predefined threshold, the image is classified as either probably benign (C3) or suspicious for malignancy (C4), accordingly. It's worth noting that this network is also built upon the pre-trained ResNET-50 model \citep{he2016deep}.

In Figure \ref{fig:network} (c), we introduce the `Probe information tool' responsible for providing information regarding the probe's position. It is tasked with categorizing probe marks in the image into 12 distinct classes, including nipple, Upper Inner Quadrant (UIQ), Upper Outer Quadrant (UOQ), Lower Outer Quadrant (LOQ), Lower Inner Quadrant (LIQ), and axillary lymph node for both left and right sides. This network is built upon the pre-trained ResNET-50 model \citep{he2016deep}. 

For comprehensive training details of these networks, please refer to Section \ref{sec:implementation}.

\section{Implementation Details}
\label{sec:implementation}

\subsection{Dataset}
To train each tool of the proposed method, we acquired BUS images about from 750 patients at Chung-Ang university hospital. Each patients contributed multiple images from various parts of breast. We subsequently divided 750 patients into two groups: 720 patients for training and 30 patients for the final report evaluation. 

For the training of the `Suspicious description tool' and the `Category classification tool', we acquired several hundred images for each class and subsequently partitioned them into training, validation, and test sets. The specific dataset sizes are provided in detail in Table \ref{table:dataset}. To train the `Probe information tool', we employed 4000, 518, and 475 images for training, validation, and testing, respectively. These images were utilized in their cropped form, focusing on the probe position mark. During the training process, we implemented various data augmentation techniques, including flipping, rotating, sharpening, and others, to address concerns related to over-fitting and data scarcity. It's important to note that all the images used for training, validation, and testing were sourced from the initial group of 750 patients.

In the final phase of report generation, we chose images from an additional set of 30 patients. For each patient, we generated preliminary reports for representative images, instructing the LLM as follows: ``Please provide a summary of the following observations: \textit{\{whole preliminary reports\}}."

\begin{table}[h!]
    \centering
    \caption{Dataset composition}
    \vspace{0.3cm}
	\resizebox{0.5\textwidth}{!}{
		\begin{tabular}{c||c|c|c|c}
			\hline 
			\multirow{2}{*}{Phase} & Shape & Margin & Echo & Category \\ 
            & (a/b) & (c/d/e) & (f/g/h) & (i/j/k) \\ \hline
			$Training$ & (213/220) & (216/130/182) & (109/86/222) & (523/247/400)\\
			$Validation$ & (27/27) & (27/15/23) & (14/11/29) & (72/46/50) \\
			$Test$ & (27/27) & (27/15/23) & (14/11/29) & (72/46/50)\\
			\hline
		\end{tabular}
	}
    \begin{tablenotes}
        \small
        \item a: irregular, b: oval, 
        \item c: circumscribed, d: indistinct, e: microlobulated
        \item f: anechoic, g: isoechoic, h: hypoechoic
        \item i: benign, j: malignant, k: normal
    \end{tablenotes}
	\label{table:dataset}
\end{table}

\subsection{Model training}
As mentioned earlier, the entire classification network is built upon the pre-trained ResNET-50 model \citep{he2016deep}. We modified the size of the final fully-connected layer to accommodate the specific number of classes for each task. During the training process, we utilized the Adam optimizer with a learning rate of $1e-4$. The network underwent training for 200 epochs, employing a batch size of 2, with the objective of minimizing the cross-entropy loss between the network's output and the ground truth labels. We selected the model that demonstrated the highest validation accuracy.

It's worth noting that we conducted our training using the PyTorch framework on a GeForce RTX 3090.

\section{Experimental Results}
\label{sec:results}

\subsection{Results for each tool}

\paragraph{Suspicious description tool}
The `Suspicious description tool' consists of three classification networks, focusing on shape, margin, and echo characteristics. Leveraging the information from these three networks, it provides a description of suspicious findings. Representative results are visualized in Figure \ref{fig:result_description} (a) to (c), while their corresponding confusion matrices are displayed in (d) to (f). As the figures illustrate, each network effectively classifies its respective categories. Notably, the magnified region within the blue box emphasizes the key areas of interest. Examining the confusion matrices, the diagonal values indicate correct classifications, with the numbers within each box representing the count of correct cases. It is evident that the network excels in classifying shape, margin, and echo characteristics accurately. This robust performance is further supported by the network's high accuracy, precision, recall, and F1 score, demonstrating the effectiveness of our custom-trained model.

\paragraph{Category classification tool}
The primary function of `Category classification tool' is to determine the category of the main image. This network classifies the main image into three classes: benign (C2), highly suggestive of malignancy (C5), and negative (C1). Notably, each category is further divided into `C2' / `C3' and `C4' / `C5' based on specific threshold probabilities. In this study, we have set the threshold values at 0.85 and 0.95, respectively.

\begin{figure}[t!]
  \center
	\includegraphics[width=0.5\textwidth]{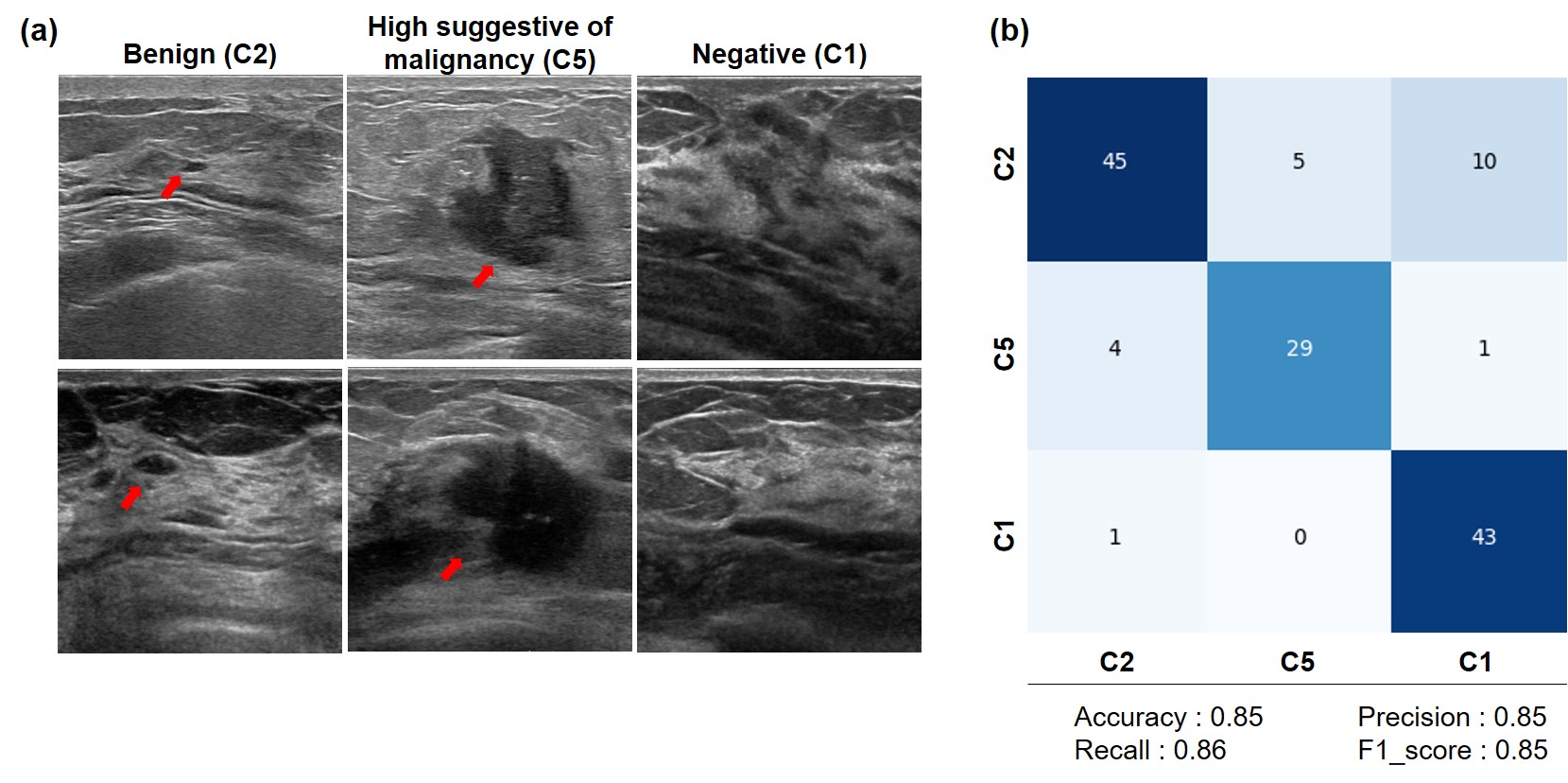}
	\vspace*{-0.3cm}
	\caption{Results of `Category classification tool'. (a) The representative results. The red arrow denotes the region to focus on. (b) The confusion matrix.}
	\label{fig:result_category}
\end{figure}

The results are graphically presented in Figure \ref{fig:result_category}. In  Figure  \ref{fig:result_category}(a), representative outcomes are displayed, showing the network's effective classification into the corresponding categories. The red arrow highlights key focal points. In  Figure  \ref{fig:result_category}(b), the confusion matrix reveals instances of mis-classification, but it's evident that the network adeptly identifies suspicious findings. The accuracy and F1 score, measuring 0.85 and 0.85, respectively, demonstrate the strong classification performance.

\begin{figure}[t!]
  \center
	\includegraphics[width=0.5\textwidth]{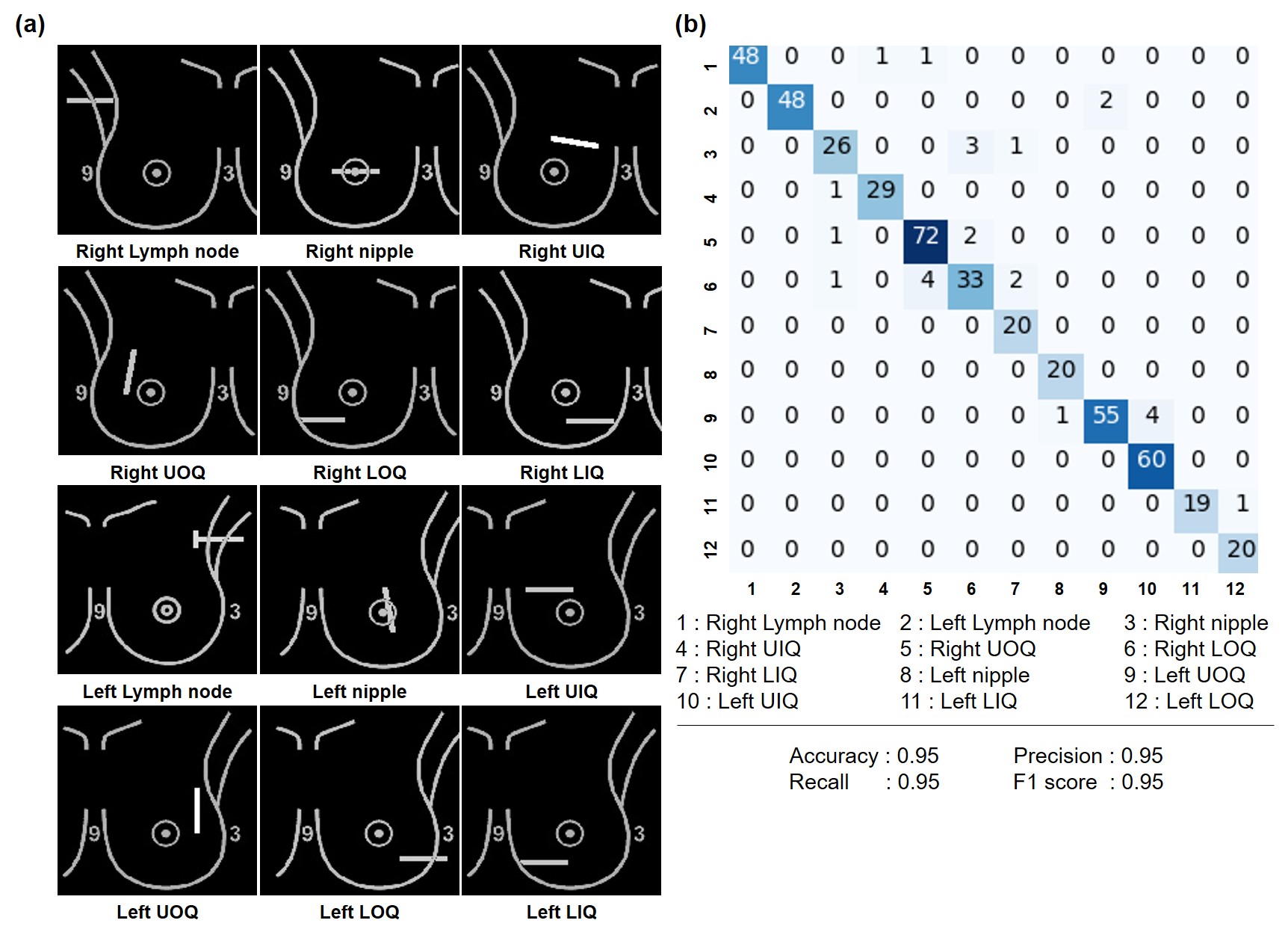}
	\vspace*{-0.3cm}
	\caption{Results of `Probe information tool'. (a) The representative results. (b) The confusion matrix.}
	\label{fig:result_probe}
\end{figure}

\begin{figure}[t!]
  \center
	\includegraphics[width=0.5\textwidth]{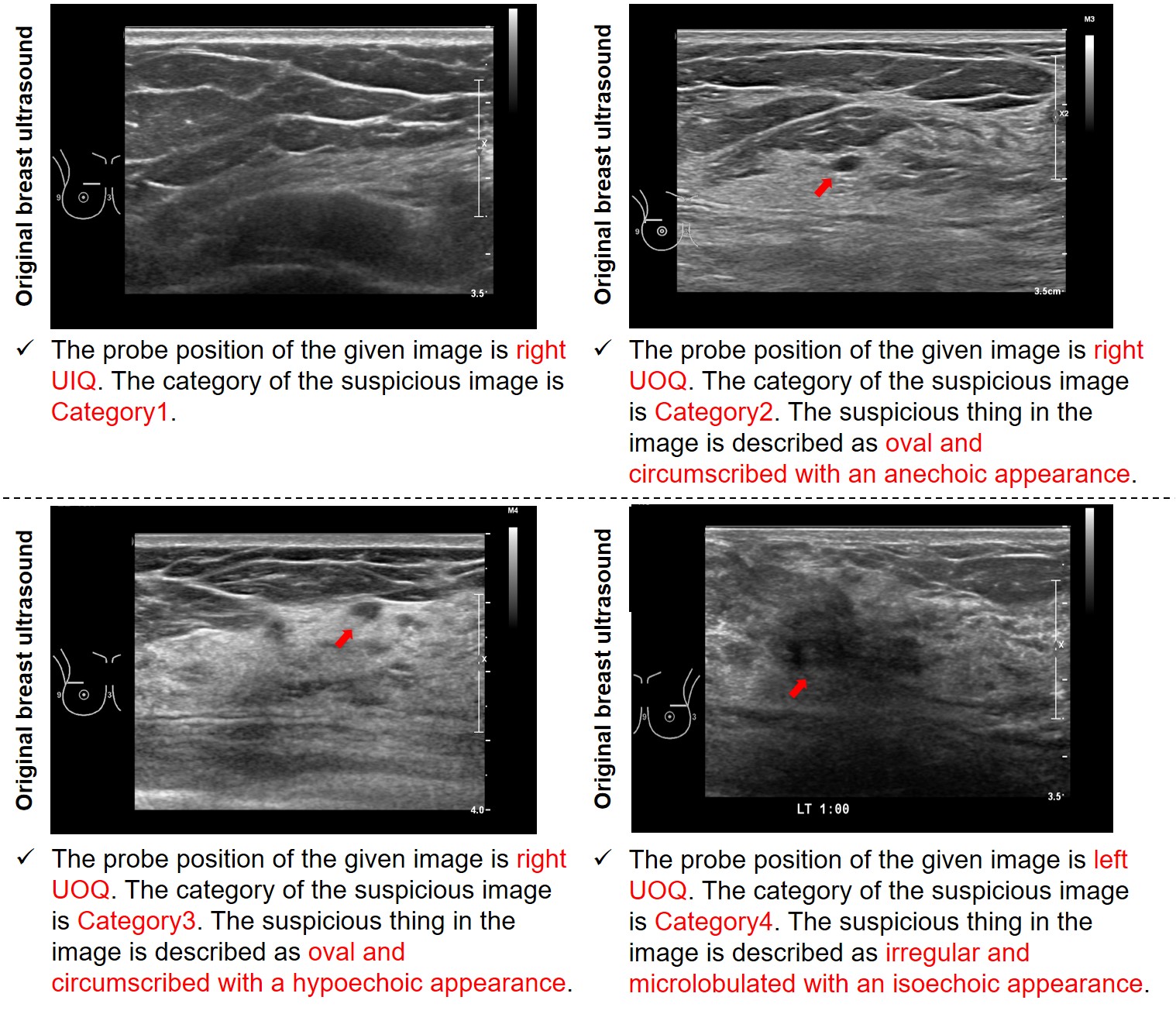}
		\vspace*{-0.3cm}
	\caption{
	Representative results of our preliminary report generation. The red denotes the generated information from each designated tools.}
	\label{fig:result_report}
\end{figure}

\paragraph{Probe information tool}
The `Probe information tool' is responsible for categorizing probe position marks into 12 distinct classes, which include right lymph node, right nipple, right UIQ, right UOQ, right LOQ, right LIQ, left lymph node, left nipple, left UIQ, left UOQ, left LOQ, and left LIQ. We assessed the quality of probe mark classification using a test set of 475 probe marks.

As illustrated in Figure \ref{fig:result_probe}(a), the network adeptly classifies the marks into their respective categories.      Figure \ref{fig:result_probe}(b) presents a confusion matrix where the darker color signifies the highest predicted count for each class. Notably, the dark color predominantly aligns with the diagonal axis, indicating the network's accurate mark classification. Furthermore, both the accuracy and F1 score stand at 0.95, supporting the network's high-quality performance.

\subsection {Preliminary report generation}
The proposed method is founded on LangChain, which features a memory system capable of retaining information from previous observations. When a prompt is introduced into LangChain, it schedules the process based on the provided instructions and the stored memory data. For instance, if we input a request like ``Give me a probe information for the given image: \textit{\{image path\}}. Then, provide a category of image and description of it." LangChain systematically engages the `Probe information tool', `Category classification tool', and `Suspicious description tool' in sequence. It records these observations and provides them in the final result, referred to as the preliminary report.

The outcomes are presented in Figure \ref{fig:result_report}, where we have included four representative results. The red annotations signify the information generated by each of the tools, encompassing probe information, image category, and details pertaining to shape, margin, and echo. It is evident that the generated preliminary reports effectively encapsulate crucial information corresponding to the respective BUS images.

\subsection{Final report generation}
\begin{figure*}[t!]
  \center
	\includegraphics[width=\textwidth]{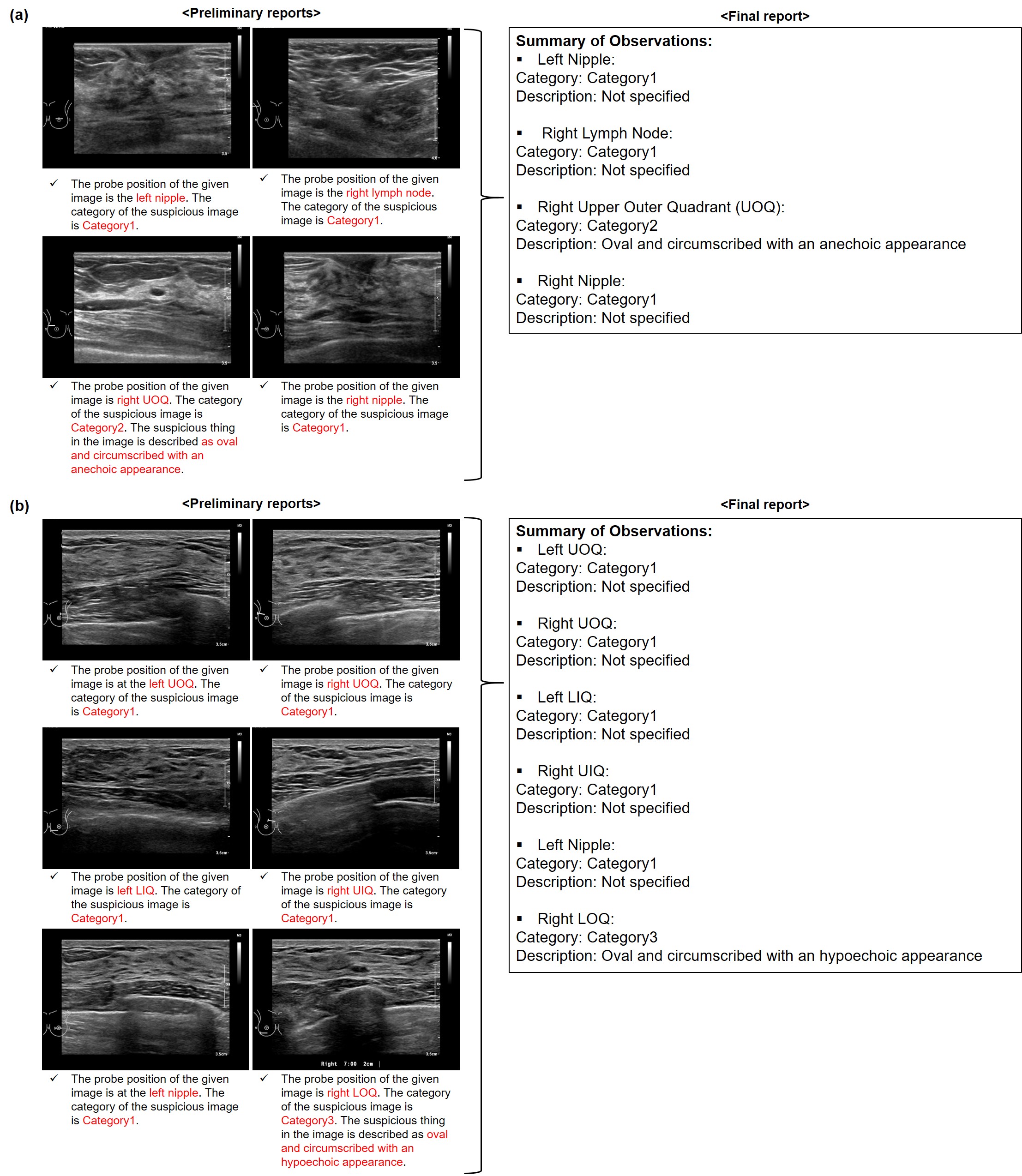}
	\caption{Final reports generated from our method. The final report is generated from preliminary report represented at left part. The red text signifies the information generated by our designated tools.}
	\label{fig:result_final}
    \vspace{0.5cm}
\end{figure*}

The final report must encompass information derived from various segments of the BUS image. To achieve this, we aggregate the preliminary reports for each BUS image and input them into a LLM with the instruction, ``Please give me a summary of the given observations depending on a probe position: \textit{\{entire preliminary report\}}." Subsequently, the LLM generates a summary from the whole preliminary report, and the results are displayed in Figure \ref{fig:result_final}.

The final report divides the information based on the probe position region, offering the category and description for each. As depicted in Figure \ref{fig:result_final}, the final report effectively integrates each part of the preliminary report and structures the information into a predefined format.

\subsection{Final report evaluation}
The resulting final report is expected to encompass all the critical information pertaining to the provided image. To assess this, we conducted evaluations with board-certified radiologists, considering three key aspects. The first is the correctness of the assigned category. The second focuses on the accuracy of the probe position. The third pertains to the precision of the suspicious thing description. By incorporating these criteria, the final report is rated on a five-degree scale, ranging from ``unacceptable" to ``poor," ``acceptable," ``good," and ``excellent." The clinical evaluation criteria are detailed in Table \ref{table:criteria}.

\begin{table}[h!]
    \centering
    \caption{Clinical Evaluation Criteria}
    \vspace{0.3cm}
	\resizebox{0.25\textwidth}{!}{
		\begin{tabular}{c|c}
			\hline 
			Score & Strength of agreement \\ \hline
			$1$ & unacceptable \\
			$2$ & poor \\
			$3$ & acceptable \\
			$4$ & good \\
			$5$ & excellent \\ \hline
		\end{tabular}
	}
	\label{table:criteria}
\end{table}

We evaluated our method using a dataset of 30 patients, with each person having a set of 5 to 15 images. The proposed method achieved an average score of 3.53, signifies that the final reports effectively capture critical information for each part.

\subsection{Adding suspicious detection tool}
To verify the comprehensiveness of our proposed method, we have integrated an additional tool for detecting suspicious elements within the BUS images. This tool determines where the suspicious thing is. These tools have been incorporated into LangChain, and the instruction provided is: ``Is there any suspicious thing in the given image? \textit{\{image path\}}."

% This tool determines whether there is anything noteworthy to detect. If nothing is found, it responds with "There is no suspicious thing." However, if something is detected, it saves the image with a bounding box.

The underlying algorithm for this network is the `PIX2SEQ' model \citep{chen2021pix2seq}, which generates bounding box indexes sequentially through next token prediction process. To train the `Suspicious detection tool', we employed 8000 images for training and 160,158 images for validation and testing. The entire network was trained over 200 epochs, using a batch size of 16. The vision encoder in the pre-trained model was initialized with weights from ImageNet, and the network was trained with a learning rate of $1e-4$ to optimize the cross-entropy loss.

% The bounding box indexes are structured in the format $(class, x_{min}, y_{min}, x_{max}, y_{max})$, with the class indicating the presence (1) or absence (0) of a suspicious element. 

\begin{figure}[t!]
  \center
	\includegraphics[width=0.45\textwidth]{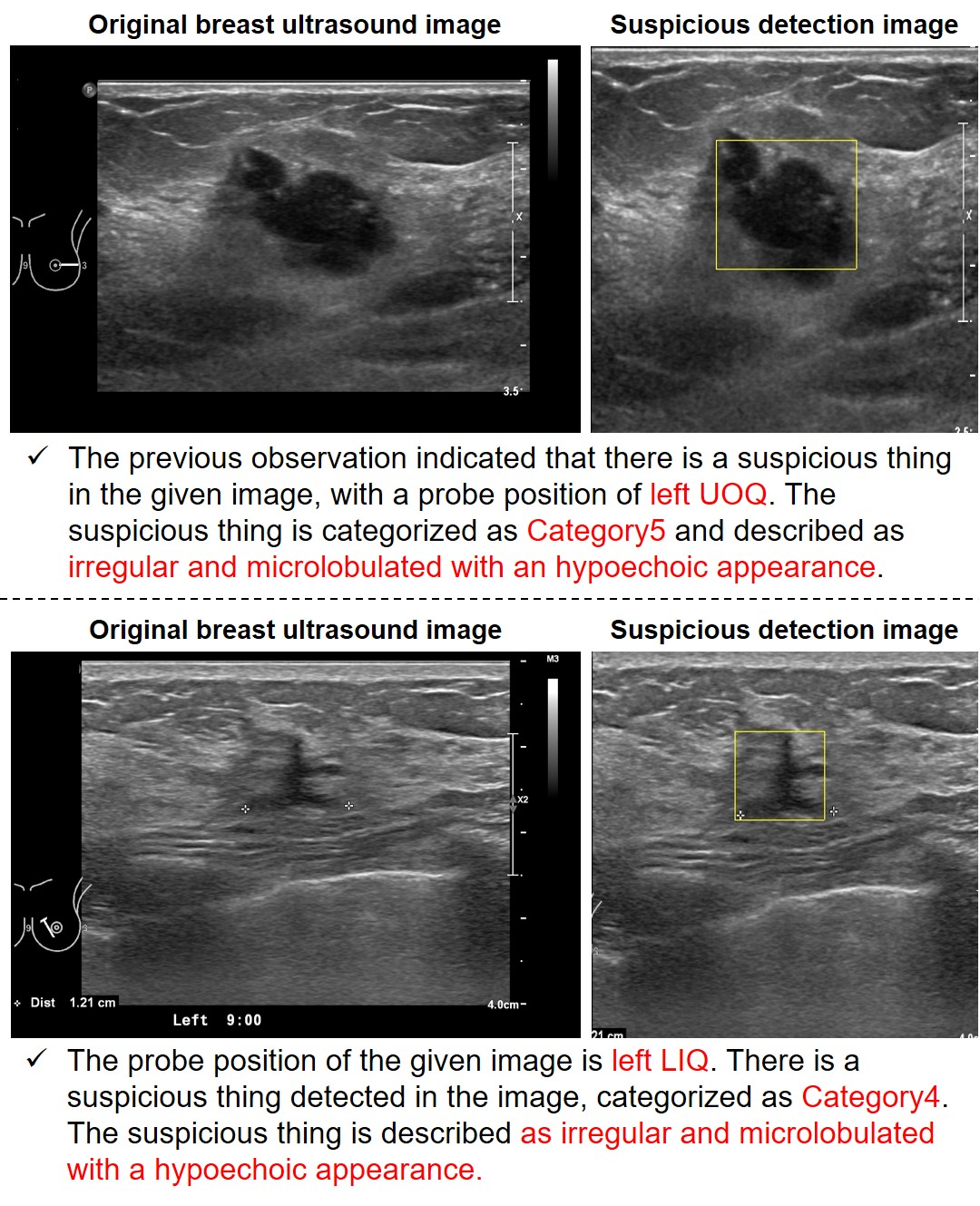}
	\vspace*{-0.3cm}
	\caption{Results of proposed method with `Suspicious detection tool'. The yellow box denotes the detected region. The red text represents the generated information from our method.}
	\label{fig:result_detection}
\end{figure}

Among the 158 test images featuring suspicious elements, we assessed the Intersection over Union (IoU) value, achieving a score of 0.684. The bounding box outcomes are shown in Figure \ref{fig:result_detection}, with the highlighted yellow box indicating the identified region via our network. These results serve as a demonstration of the network's precise identification of suspicious elements.

\subsection{Adding Optical Character Recognition (OCR) tool}
\begin{figure}[t!]
  \center
	\includegraphics[width=0.5\textwidth]{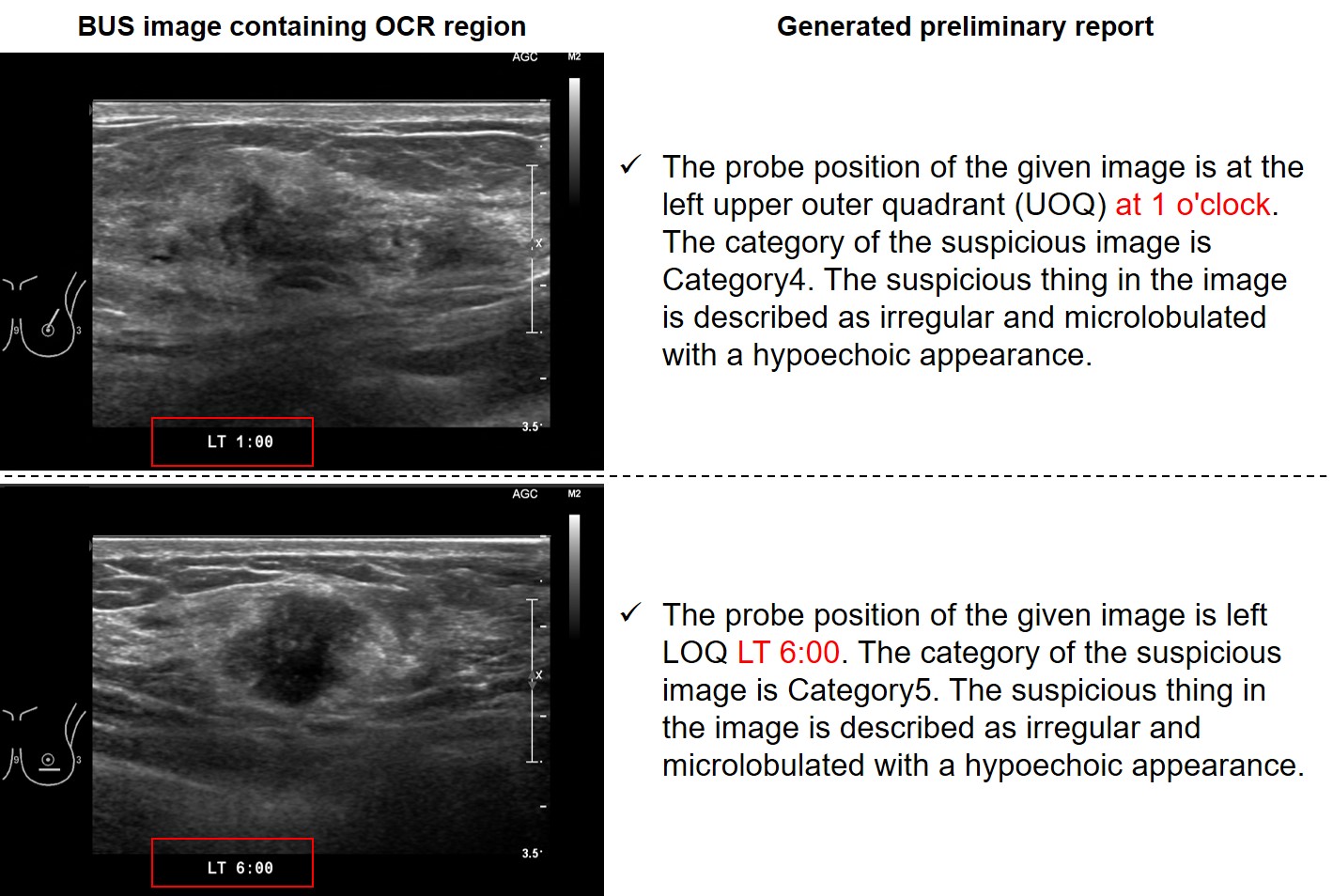}
	\vspace*{-0.3cm}
	\caption{Results of the proposed method with `OCR tool'. The red box denotes the region to recognize. The red text represents the generated information from combination of `Probe information tool' and `OCR tool'.}
	\label{fig:result_ocr}
\end{figure}

To assess the scalability of the proposed method, we incorporated an additional tool designed to identify the probe's specific information within the image. This tool functions as an Optical Character Recognition (OCR) model crucial for interpreting nuanced image specifics, including the probe's distance from reference points and directional details. To implement this aspect, we utilized a pre-trained model sourced from TrOCR \citep{li2023trocr}, which is based on transformer architecture. The outcomes are illustrated in Figure \ref{fig:result_ocr}. In the visual representation, the red box delineates the OCR region, and in the preliminary report, information highlighted in red signifies the output derived from the OCR tool.

\section{Discussion and Conclusions}
\label{sec:Discussion}

AI-based applications are rapidly emerging, driven by advancements in techniques and hardware. Among these applications, image diagnosis and description are particularly in demand and have been the focus of extensive research. Traditionally, radiologists manually describe each image, a time-consuming task. However, recent developments in deep learning techniques have automated this process, significantly improving clinician workflow.

The conventional report generation model, based on CNN-LSTM, has performance limitations and is unsuitable for complex images like ultrasound DICOM images. The introduction of LLM, which excel in NLP tasks involving logical reasoning, has been a game-changer. Utilizing LLM as a manager for coordinating various AI models addresses several of LLM's limitations, especially when dealing with tasks that require sub-tasks or a multi-modal approach.

Building on this pioneering work, we present a novel BUS report generation method based on LangChain, harnessing the logical reasoning capabilities of LLM. Our model employs LLM as a manager to coordinate various AI pre-trained models, providing their outputs in text format. It comprises five trained models for analyzing individual BUS images. Leveraging LangChain's memory system, our model can synthesize and deliver comprehensive reports based on prior observations. Our experiments have demonstrated that our sub-models yield satisfactory results in describing BUS images, with the preliminary reports containing essential diagnostic information. Notably, our method allows for the generation of final reports for individual patients by summarizing all their images.

Nonetheless, our study has certain limitations. First, each sub-model can introduce errors, and as a result, a single report may accumulate errors from each sub-model. Second, the image splitting process is rule-based, and it may not handle cases where the positions of probe marks or main images differ significantly. 

In summary, we have presented a BUS report generation method that assembles preliminary reports from each segment, aggregating results from sub-models to analyze images and generate comprehensive reports. These sub-models can be readily replaced or supplemented with models of similar purpose if they exhibit enhanced performance. We are confident that our method holds considerable promise for integration into medical workflows.

\section*{Acknowledgement}
This work was supported by the National Research Foundation of Korea under Grant NRF-2020R1A2B5B03001980.

\bibliographystyle{model2-names.bst}\biboptions{authoryear}
\bibliography{ref}

\end{document}